\documentstyle[12pt]{article}
\input{epsf}
\begin{document}
%
\newcommand{\nc}{\newcommand}
\def\veffT{\Delta V_{{\rm eff},T}}
\def\msh{m^2_{h^0}}
\def\veffo{V_{{\rm eff},0}}
\nc{\beq}{\begin{equation}}
\nc{\eeq}{\end{equation}}
\nc{\beqa}{\begin{eqnarray}}
\nc{\eeqa}{\end{eqnarray}}
\nc{\lra}{\leftrightarrow}
\nc{\sss}{\scriptscriptstyle}
{\nc{\lsim}{\mbox{\raisebox{-.6ex}{~$\stackrel{<}{\sim}$~}}}
{\nc{\gsim}{\mbox{\raisebox{-.6ex}{~$\stackrel{>}{\sim}$~}}}
\def\dsl{\partial\!\!\!/}
\def\lameff{\lambda_{\rm eff}}
\def\Re{{\rm Re\,}}
\def\Im{{\rm Im\,}}
\begin{titlepage}
\pagestyle{empty}
\baselineskip=21pt
\rightline{McGill 96-16}
\rightline{hep-ph/9609240}
\rightline{September 3, 1996}
\vskip .4in
\begin{center}
{\large{\bf Electroweak Phase Transition \\
in Two Higgs Doublet Models}}
\end{center}
\vskip .1in
\begin{center}
James M.~Cline\\
and\\
Pierre-Anthony Lemieux\footnote{Present address: Department of Physics,
UCLA, 405 Hilgard Ave., Los Angeles, California 90095-1547}\\
{\it McGill University, Montr\'eal, Qu\'ebec H3A 2T8, Canada}
\vskip .2in
\end{center}

\vskip 0.7in \centerline{ {\bf Abstract} } 
\baselineskip=18pt 
\vskip
0.5truecm 

We reexamine the strength of the first order phase transition in the
electroweak theory supplemented by an extra Higgs doublet.  The
finite-temperature effective potential, $V_{\rm eff}$, is computed to
one-loop order, including the summation of ring diagrams, to study the
ratio $\phi_c/T_c$ of the Higgs field VEV to the critical temperature.
We make a number of improvements over previous treatments, including a
consistent treatment of Goldstone bosons in $V_{\rm eff}$, an accurate
analytic approximation to $V_{\rm eff}$ valid for any
mass-to-temperature ratios, and use of the experimentally measured top
quark mass.  For two-Higgs doublet models, we identify a significant
region of parameter space where $\phi_c/T_c$ is large enough for
electroweak baryogenesis, and we argue that this identification should
persist even at higher orders in perturbation theory.  In the case of
the minimal supersymmetric standard model, our results indicate that
the extra Higgs bosons have little effect on the strength of the phase
transition.

\end{titlepage} 
\newpage

\baselineskip=20pt

\section{Introduction}

There is now convincing evidence that the baryon asymmetry of the
universe could have been created at the electroweak phase transition
(EWPT) \cite{Shap}, with a minimal amount of new, yet plausible physics
beyond the standard model \cite{NKC}-\cite{CKV}.  However for
electroweak baryogenesis to work, it is necessary that the
baryon-violating interactions induced by electroweak sphalerons be
sufficiently slow immediately after the phase transition to avoid the
destruction of the baryons that have just been created.  This condition
is fulfilled if the vacuum expectation value (VEV) of the Higgs field
in the broken phase is large enough compared to the critical
temperature at the time of the transition (see for example
ref.~\cite{reviews} and references therein),
\beq
        \phi_c/ T_c   > 1.
\label{cond} 
\eeq 
In the standard model, the bound (\ref{cond}) holds only for very light
Higgs bosons \cite{MC}-\cite{DLHLL} already ruled out by experiment,
$m_{h^0}< 20$ GeV, according to a recent nonperturbative study of the
phase transition \cite{KLRS1}.  In addition to the difficulties with
producing a large enough initial baryon asymmetry, the impossibility of
satisfying the sphaleron constraint (\ref{cond}) in the standard model
provides an incentive for seeing whether the situation improves in
extended theories.

The EWPT in two-Higgs doublet models has been previously studied in
references \cite{BKS}-\cite{DFJM}.  In the present work we improve on
the approximations made in the previous studies in the following
ways.\footnote{The list is not meant to imply that all of the previous
studies were subject to all of the limitations mentioned here.  For
example, only ref.~\cite{TZ} used unitary gauge.}

\noindent$\bullet$
Avoidance of the unitary gauge, which is known to be problematic: it has
been shown that it is necessary to resum an infinite number of contributions
in the perturbative expansion to make the determination of finite-temperature
quantities agree between unitary gauge and renormalizable gauges \cite{ABV}. 

\noindent$\bullet$ Inclusion of the ${\mu_3^2} \Phi_1^\dagger\Phi_2$ term
in the Higgs potential necessary for CP violation, which is crucial for
the baryon asymmetry production mechanism.  In contrast to
ref.~\cite{DFJM}, we find that the strength of the transition does not
necessarily grow with increasing ${\mu_3^2}$.

\noindent$\bullet$  Use of the ring summation
\cite{MC}-\cite{AE},\cite{Par} for all the bosonic thermal loop
contributions to the effective potential.  Furthermore we compare two
different prescriptions for implementing ring-improvement.  The
differences are indicative of the theoretical uncertainties inherent in
using one-loop perturbation theory.

\noindent$\bullet$ Use of a critical temperature $T_c$ more closely
corresponding to the onset of bubble nucleation, namely the temperature
where the two minima of the potential are degenerate, rather than that
where the second derivative of the potential vanishes at the origin of
field space.  We find a quantitative difference between the two methods
when determining whether a given set of parameters in the model is
really consistent with the sphaleron bound.

\noindent$\bullet$ Use of the experimentally measured top quark mass,
$175\pm 6$ GeV \cite{topmass}, which was not available when the
previous papers \cite{BKS}-\cite{DFJM} were written.  The large mass of
the top quark greatly reduces the strength of the phase transition in
the standard model \cite{AE}-\cite{DLHLL}.

\noindent$\bullet$ Estimating the expansion parameter which quantifies
the convergence of the perturbative expansion at the phase transition,
near the critical value of the Higgs field VEV\cite{SW,AE}.  This is
the major potential weakness of the perturbative approach being used
here.  However we find that according to this criterion, most of the
promising regions of parameter space are relatively safe from the
infrared divergences that would cast doubt on the convergence of the
loop expansion.

\noindent$\bullet$ Inclusion of the Goldstone boson contributions to
the effective potential.  These can become important when the lightest
Higgs boson $h^0$ starts to become as heavy as the other particles in
the theory.  Although the case of large $m_{h^0}$ is uninteresting in
the standard model, from the point of view of having a strongly first
order phase transition, in extended models the effect of large
$m_{h^0}$ can be compensated by new particles.

\noindent$\bullet$ Avoiding the expansion in mass over temperature that
would limit the scope of our investigation.  This turns out to be quite
important since the region of parameter space where the phase
transition is strongly first order is largely where the conventional
expansion of the effective potential in powers of $M/T$ is starting to
fail badly.

A price we pay for these improvements is the restriction of the
potential such that the ratio of the two Higgs VEV's ($\tan\beta$) is
unity, to make the problem more tractable; thus we are not exploring
the entire parameter space of the class of models of interest.  On the
other hand, this reduction has the advantage of making it easier to
characterize the favored values of the parameters in the Higgs
potential.  Furthermore a number of studies of the minimal
supersymmetric standard model (MSSM) indicate that the strength of the
phase transition is generally greatest for small values of $\tan\beta$
\cite{eqz}-\cite{laine}. In section 4 we discuss the significance of our
results for the MSSM in the region of $\tan\beta=1$.

\section{Effective Potential}
\subsection{Zero temperature; Goldstone boson contributions}

We begin by constructing the effective potential for the CP-even
component of the Higgs field that is responsible for electroweak
symmetry breaking.  At one loop and at zero temperature, each particle
makes a Coleman-Weinberg-type contribution to $\veffo(\phi)$
\cite{CW}.  It can be written in the form of a supertrace\footnote{each
real scalar field has a weight of $+1$ in the supertrace, and each
Dirac fermion has a weight of $-4$} involving the matrix $M$ of all the
field-dependent particle masses,

\beq
        \veffo(\phi) = \frac{\lambda}{4} (\phi^2 - v^2)^2 +\frac12 A\phi^2
         +\frac{1}{64\pi^2}{\rm Str}\,
         M^4(\phi) \left(\log {M^2(\phi)\over \mu^2} - \frac32\right),
\label{zeroT} 
\eeq
where the pieces proportional to $A$ and $\log(\mu^2)$ are counterterms
that must be fixed by renormalization conditions.   A convenient choice
of the latter is that the tree-level definitions of the VEV ($v=246$ GeV})
and the Higgs mass $m^2_{h^0} = V''(v) = 2\lambda v^2$ should be
maintained.  These conditions are given by\baselineskip=20pt
\beqa
	\label{vevcond}
	0 &= V'(v) &= A v +  \frac{1}{32\pi^2}{\rm Str}\,M^2 {M^2}'
	\left(\log {M^2\over \mu^2} - 1\right)\\
	\label{masscond}
	m^2_{h^0} &= V''(v)  &= 2 \lambda v^2 + A + 
 	\frac{1}{32\pi^2}{\rm Str}\left[\left(M^2 {M^2}''+ ({M^2}')^2\right)
	\left(\log {M^2\over \mu^2} - 1\right) + ({M^2}')^2\right]
\eeqa\baselineskip=20pt
where the prime denotes $\partial/\partial\phi$ and all masses are 
evaluated at $\phi=v$.  One can always choose $A$ such that 
eq.~(\ref{masscond}) is satisfied when $m^2_{h^0}=2 \lambda v^2$, so the
tree-level relation between $m^2_{h^0}$ and the VEV is preserved. 
\baselineskip=20pt

However there is a technical problem with eq.~(\ref{masscond}): in the
Landau gauge which is the most convenient one because of the decoupling
of the ghosts, the Goldstone bosons are massless at $\phi=v$, but they
have a nonvanishing value of ${M^2}'(v)$, giving a logarithmic infrared
divergence due to the term $({M^2}')^2\log {M^2}$, which does not occur in
other gauges.  Yet physical masses must be gauge-invariant and
IR-finite.  The problem is that $m^2_{h^0}$ as defined in
(\ref{masscond}) is not the physical pole mass, which must be evaluated at
an external momentum of $p^2=m^2_{h^0}$, but rather the off-shell
self-energy evaluated at $p^2=0$.  To avoid the IR divergence and
the gauge dependence we should really evaluate the pole mass.  By
computing the Feynman diagrams of the type shown in figure 1 which are
responsible for the IR problem, it is easy to see how to fix it: 
in eq.~(\ref{masscond}) one should make the replacement
\begin{figure}[b]
\epsfysize=0.5in\epsfbox{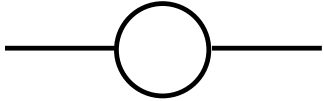} 
\caption{self-energy diagram responsible for IR-divergent contribution
of Goldstone bosons to the Higgs boson mass at $p^2=0$, in Landau gauge.}
\label{fig:loop}
\end{figure}
\beqa
	\label{presc}
	\log M^2 &\to& {\rm Re }\,\int_0^1 dx \log \left( M^2 - x(1-x)
	m^2_{h^0}\right) \nonumber\\
	&=& \log m^2_{h^0} - 2 + \left\{ \begin{array}{ll}
	\sum_\pm (1 \pm\sqrt{1-4R})\log(\frac12 \pm\sqrt{\frac14-R}),
	& R<\frac14;\\
	\log R + 2\sqrt{4R-1}\,\tan^{-1}\left({1\over\sqrt{4R-1}}\right),&
	R>\frac14; \end{array} \right.\\
	&\equiv& F(M^2, m^2_{h^0}); \qquad R\equiv M^2/m^2_{h^0},\nonumber
\eeqa
Using the prescription of
eqs.~(\ref{masscond}-\ref{presc}), and the convention $m^2_{h^0}=2
\lambda v^2$, it is straightforward to solve for the renormalization
constants:
\beqa
	\log\mu^2 &=& {\rm Str}\left[\left({M^2}{M^2}'/v + ({M^2}')^2
	\right)  F(M^2, m^2_{h^0}) - ({M^2}{M^2}'/v)\log(M^2)\right]
	/  {\rm Str}\, ({M^2}')^2;\nonumber\\
	A &=& -\frac{1}{32\pi^2 v }{\rm Str}\, M^2 {M^2}'\left(
	\log {M^2\over\mu^2} - 1\right).
\eeqa 
We used the fact that for any renormalizable theory,  ${M^2}'' =
{M^2}'/v$, to simplify the result.

By this procedure we are able to consistently include the contributions
of the Goldstone bosons into the effective potential (\ref{zeroT}).  In
the standard model, for small values of $m^2_{h^0}$ which were of
interest for maximizing the strength of the phase transition, both the
usual Higgs boson and thus the Goldstone bosons make numerically small
contributions to the effective potential and can thus be safely
ignored.  In two Higgs doublet models however, we will show that it
is possible to get strong phase transitions even when $m^2_{h^0}$
becomes rather large, and therefore the Goldstone boson contributions
can be nonnegligible.

\subsection{Finite temperature}

At one-loop order but now at finite temperature, there is an additional
contribution to the effective potential given by \cite{DJ}
\beq
        \veffT(\phi) = \frac{T^4}{2\pi^2}{\rm Str}\int_0^\infty
        dx\,x^2\ln\left(1\mp e^{-\sqrt{x^2 + M^2(\phi)/T^2}}\right)
\label{finiteT} 
\eeq 
where the sign is $-$ for bosons and $+$ for fermions.  For small $M/T$
the contribution per degree of freedom can be expanded as\footnote{see
eq.~(3.32) of ref.~\cite{AE}}
\beqa
&V_{\rm s,b}(n) = -\frac{\pi^2 T^4}{90} + 
\frac{M^2 T^2}{24} - \frac{M^3 T}{12\pi} - \frac{M^4}{64 \pi^2}
\left(\log\left(\frac{M^2}{T^2}\right) - c_b\right)& \nonumber\\
&+ \frac{M^2 T^2}{2}{\displaystyle\sum^n_{l=2}} 
\left(\frac{-M^2}{4\pi^2 T^2}\right)^l
\frac{(2l-3)!! \zeta(2l-1)}{(2l)!!(l+1)}, &{\rm bosons};\nonumber\\
&V_{\rm s,f}(n) = -\frac{7\pi^2 T^4}{720}+\frac{M^2 T^2}{48}  + 
\frac{M^4}{64 \pi^2}
\left(\log\left(\frac{M^2}{T^2}\right) - c_f\right)\nonumber\\
&- \frac{M^2 T^2}{2}{\displaystyle\sum^n_{l=2}}
 \left(\frac{-M^2}{4\pi^2 T^2}\right)^l
\frac{(2l-3)!! \zeta(2l-1)}{(2l)!!(l+1)}\left(2^{2l-1} - 1\right)
,  &{\rm fermions};\nonumber\\
&c_b = 3/2 + 2\log 4\pi - 2\gamma_{\sss E} \cong 5.40762;\qquad
c_f = c_b -2\log4 \cong 2.63503& 
\label{small}
\eeqa	 
where $\zeta$ is the Riemann $\zeta$ function.
The term cubic in the boson masses is especially important because it
gives rise to the barrier in the potential that makes the transition first
order, as needed for baryogenesis.  For large $M/T$ on the other hand,
the contribution from either bosons or fermions has the asymptotic
expansion\footnote{see eq.~(B5) of ref.~\cite{AH}}
\beq
V_{\rm l}(n) = 	-  e^{-M/T}\left(\frac{ M T }{2\pi}\right)^{3/2}
\sum_{l=0}^n \frac{1}{2^l l!}\,\frac{\Gamma(5/2+l)}{\Gamma(5/2-l)}
(T/M)^l.
\label{large}
\eeq

For the purpose of numerical analysis of the phase transition, it is
useful to have an analytic approximation to the exact expression
(\ref{finiteT}) since it is computationally expensive to evaluate the
integral.  However, by smoothly joining the small $M/T$ expansion
(\ref{small}) with that for large $M/T$, eq.~(\ref{large}), one can
obtain an excellent approximation to the exact integral.  To do this
optimally, one should choose the value of $M/T$ where the derivatives
of the two expansions match each other as the transition point.  Using
$V_{\rm l}(3)$ for the large $M/T$ approximation, we found good matching
to the exact integral by using
$V_{\rm s,b}(3)$ for bosons and $V_{\rm s,f}(5)$ for fermions.
This gives an approximation with a relative error which is less
than 0.5\% for $M/T = \infty$, and negligible for small $M/T$.
Specifically, we took
\beqa 
\veffT = {\rm Tr} \Bigl[
 & \left[\Theta(x_b-(M/T)^2)\, V_{\rm s,b}(3) + 
	\Theta((M/T)^2-x_b)\,(V_{\rm l}(3)-\delta_b)\right] P_b 
	& {\rm}\nonumber\\
 & \left[	\Theta(x_f-(M/T)^2)\, V_{\rm s,f}(4) + 
	\Theta((M/T)^2-x_f)\, (V_{\rm l}(3)- \delta_f)\right] P_f 
	& {\rm }\label{matching}
	\Bigr],
\eeqa
where $\Theta$ is the step function with $x_b = 9.47134$ and $x_f = 
5.47281$, and $P_{b,f}$ denotes the projection operator for bosons
or fermions, respectively.  The small constant shifts of $V_{\rm l}(3)$
are made so that the function as well as its derivatives match at the
transition point: $\delta_b = 3.1931\times 10^{-4}$ and $\delta_f = 
4.60156\times 10^{-4}$.

The above discussion implicitly assumed real-valued masses, but for
completeness there is one further matching of large and small $M/T$
behavior which must be considered separately: it can happen that some
of the bosons have large and negative values of $M^2$, corresponding to
imaginary $M/T$.  While it may be arguable whether eq.~(\ref{finiteT})
has a meaningful physical interpretation for negative $M^2$, to the
extent that it has any meaning, it is clearly necessary to use the
large $M/T$ expansion when $|M^2/T^2|$ is large.  For negative values
of $(M/T)^2$ the matching conditions are different; in
eq.~(\ref{matching}) one should use $x_b = -6.84368$ and $\delta_b =
0.045374$.  Since large values of $|M^2/T^2|$ always imply the
decoupling of the heavy particles, none of our results concerning the
interesting regions of parameter space, where the sphaleron constraint
eq.~(\ref{cond}) is satisfied, will depend on the precise value of the
effective potential when the squared masses are large and negative.

One can easily improve the above one-loop result by resumming a
subclass of thermal loops known as the ring diagrams, which amounts to
finding the thermal corrections to the boson masses\footnote{the reason
this does not apply to the fermion masses can be understood most easily
in the imaginary time formalism of finite-temperature field theory,
where the effective squared masses of the Matsubara modes are
$M^2(\phi,T) + (2\pi n T)^2$ for bosons and $M^2(\phi,T) + (2\pi
(n+\frac12 T)^2$ for fermions.  Only for the $n=0$ modes of the bosons
can there be an infrared divergence due to vanishing $M^2(\phi)$ which
would make it important to include the perturbative $g^2 T^2$
contribution to $M^2$.} due to $\veffT(\phi)$, typically of the form
$M^2(\phi,T) = M^2(\phi) + g^2 T^2$, and substituting them back into
the expression for the total effective potential \cite{Par}.  In fact
the choice of exactly how to resum is not unique, and Arnold and Espinosa
\cite{AE} have justified the consistency of the widely-used procedure
of replacing $M^2(\phi) \to M^2(\phi,T)$ only in the cubic term $M^3
T$ of eq.~(\ref{small}) rather than everywhere in the effective
potential.  Defining the unresummed total one-loop effective potential
as
\beq
	V_{\rm tot} = \veffo + \veffT,
\eeq
the ring-improved potential in the two methods can be written as
\beqa
 V_{\rm ring} = 
V_{\rm tot}[M^2(\phi,T)], 
&&\hbox{Parwani method}\nonumber\\
 V_{\rm ring} = V_{\rm tot}[M^2(\phi)] + 
	\frac{T}{12\pi}\sum_{\rm bosons}(M^3(\phi) - M^3(\phi,T)),
	&& \hbox{Arnold-Espinosa method}\nonumber
\eeqa
We shall investigate and compare both of these prescriptions.  Since
they differ by terms which are of two-loop order, they can give us some
idea of the uncertainties in our calculation coming from the neglect of
higher orders in perturbation theory.

The ring summation is essential for correctly estimating the
contributions from the longitudinal gauge bosons, which get a much
larger thermal mass than their transverse counterparts, because this
tends to reduce the effectiveness of the cubic term in making the phase
transition first order.  In addition, the thermal corrections decrease
the region of field space near $\phi=0$ where some of the squared
masses of the bosons become negative, giving rise to a complex
potential.  However the imaginary part of the potential can still
persist because of the genuine instability of field configurations with
small values of $\phi$.  We find that the real part of the potential
often exhibits an additional local minimum especially when some of the
bosons have a small or negative $M^2$, as illustrated in figure 2.
Such a situation indicates that the negative squared masses of the
Higgs bosons are playing a large role and that perturbation theory
should not be trusted due to infrared divergences in the effective
three dimensional theory which describes the high temperature limit.
We discuss the consequences of this phenomenon further below.

\begin{figure}[t]
\vspace{-1in}
\epsfysize=5in\epsfbox{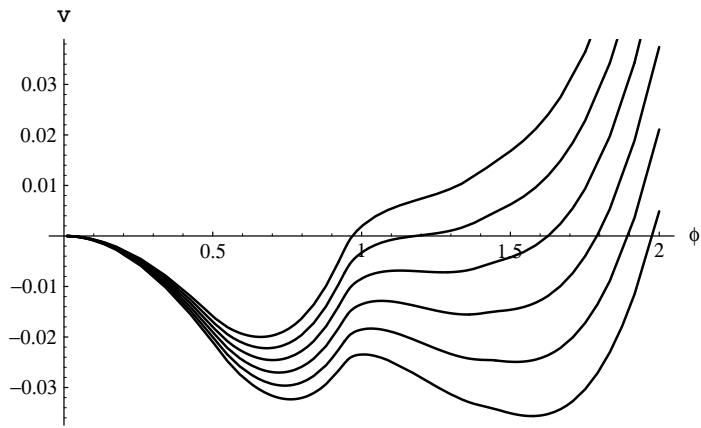} 
\vspace{-1in}
\caption{Evolution of the effective potential with temperature, showing
the pathological case of two nontrivial minima, indicative of negative 
squared boson masses.  The parameter values are $m_{h_0}=125$ GeV,
${\mu_3^2} = (30$ GeV)$^2$ and all other Higgs boson masses 187 GeV.}
\label{fig:veff}
\end{figure}

\section{Two Higgs Doublet Model}

Let us now introduce the new physics.  For the two Higgs doublet model,
including the couplings to the top quark, the potential is taken to be
\beqa
     V_{\rm 2-doublet}(\Phi_1,\Phi_2) &=&
      - \mu_1^2 \Phi_1^\dagger \Phi_1 - \mu_2^2 \Phi_2^\dagger \Phi_2
      -{\mu_3^2} \Phi_1^\dagger \Phi_2 - {\mu_3^2}^* \Phi_2^\dagger \Phi_1 +
      \frac{\lambda_1}{2} (\Phi_1^\dagger \Phi_1)^2 +
      \frac{\lambda_2}{2} (\Phi_2^\dagger \Phi_2)^2\nonumber\\
        &+& h_1(\Phi_1^\dagger \Phi_2)(\Phi_2^\dagger \Phi_1)
         + h_2(\Phi_1^\dagger \Phi_1)(\Phi_2^\dagger \Phi_2) +
      h_3\left((\Phi_1^\dagger \Phi_2)^2 + (\Phi_2^\dagger \Phi_1)^2)
      \right)\nonumber\\
         &+& \overline T_{\sss L}(y_1\Phi_1 +y_2\Phi_2)t_{\sss R} +
        {\rm\ c.c.} 
\label{twoHiggs} 
\eeqa 
It incorporates a softly-broken discrete symmetry, $\Phi_i\to -\Phi_i$,
in order to reduce the number of possible couplings.  This is the same
model as discussed in ref.~\cite{TZ}, except for the addition of the
term ${\mu_3^2} \Phi_1^\dagger \Phi_2$, which is necessary for having CP
violation in the Higgs sector.  To simplify the analysis, we follow
ref.~\cite{TZ} by setting $\mu_1^2 =\mu_2^2$, $\lambda_1 = \lambda_2$
and $y_1 = y_2$.  Furthermore we assume that ${\mu_3^2}$ is real and
positive.\footnote{The sign of ${\mu_3^2}$ is conventional since it can
always be changed by the field redefinition $\Phi_1\to -\Phi_1$.
Moreover from the results of ref.~\cite{DFJM} it appears that the
strength of the transition depends only upon $|{\mu_3^2}|$ when ${\mu_3^2}$
is complex.} This choice insures that the symmetry will break along the
direction $\Phi_1 = \Phi_2 \equiv\frac12(0, \phi)^{\sss T}$ at any
temperature.  The resulting potential for $\phi$ is identical to the
standard form $\lambda(\phi^2-v^2)^2/4$ used in (\ref{zeroT}) if one
identifies 
\beq
	\lambda = \frac14(\lambda_1+h_1+h_2+2h_3); \qquad v^2 =
	\frac{\mu^2_1 + {\mu_3^2}}{\lambda}.  
\label{lameq} 
\eeq 
The spectrum  consists of the light Higgs boson $h^0$ associated with
the field $\phi$, three Goldstone bosons $G^0, G^\pm$, the charged
particles $H^\pm$, and the neutral scalar and pseudoscalar, $H^0$ and
$A^0$ repsectively.

All that remains to complete the definition of the effective potential
is to specify the field- and temperature-dependent masses.  For the
gauge bosons these are the eigenvalues of the mass matrix in the
basis $W^\pm$, $W^3$, $B$,
\beq
        m^2_{\rm gauge}(\phi, T)={\phi^2\over 4}\left(\begin{array}{cccc}
        g^2 & & & \\  & g^2 & & \\ & & g^2 & g g' \\ & & g g' & g'^2 
        \end{array}\right) + T^2\left(\begin{array}{cccc}
        g^2 & & & \\  & g^2 & & \\ & & g^2 &  \\ & & &  g'^2 
        \end{array}\right)
        \left\{\begin{array}{rl}2, & {\rm longitudinal;}\\ 
        0, & {\rm transverse.}\end{array}\right.
\label{gauge}
\eeq
Notice that only the longitudinal gauge bosons get a thermal mass at this
order (for the transverse bosons there is a magnetic mass of order
$g^2 T$ which we neglect) and the coefficient $2$ in the
thermal correction would be $11/6$ in the absence of the second Higgs
doublet.  Also the longitudinal photon is no longer massless at finite
temperature, due to mixing between $B$ and $W^3$.  The top quark mass is
\beq
      m_{t}(\phi,T) = m_t {\phi\over v}.
\eeq
Although the dispersion relation for the top quark does get temperature
corrections which can be interpreted as a thermal contribution to its
mass squared, these do not appear in the resummation of the ring
diagrams because it is a fermion, as explained above.  For the Higgs
bosons, it is convenient to eliminate all the dimensionless couplings
and express the general masses in terms of the zero-temperature ones
and the parameter ${\mu_3^2}$ \cite{CKV}:
\beqa
        m^2_{i}(\phi,T) &=& \left\{\begin{array}{ll}
        \frac12 \left(- \msh + 
                (\msh + 2m^2_i){\phi^2\over v^2}\right) +
                \left(a+{y_1 y_2\over 4}\right)T^2, & i = h^0, G^0, G^\pm;\\ 
         \frac12 \left(- \msh +4{\mu_3^2} + 
                (\msh -4{\mu_3^2} + 2m^2_i){\phi^2\over v^2}\right) +
               \left(a-{y_1 y_2\over 4}\right)T^2, & i = A^0, H^0, H^\pm;
        \end{array}\right.\nonumber\\
        a &=& {\lambda\over 2}+\frac{3g^2+g'^2}{16}+\frac{y_1^2+y_2^2}{8}
                +\frac{1}{12} \left\{4\lambda + {1\over v^2}
                (-8{\mu_3^2} + 2m^2_{H^\pm} + 
                m^2_{H^0} + m^2_{A^0}) 
                \right\};\nonumber\\
\label{Higgsmasses}
\eeqa
For comparison, the standard model can be recovered by ignoring the
masses of $A^0, H^0, H^\pm$ and the term in curly brackets in $a$, and
setting $y_2 = 0$.  The particle multiplicities are 1 for each of the 
Higgs bosons and the longitudinal gauge bosons, 2 for each transverse
gauge boson, and 12 for the top quark.

\section{Phase Transition Results}

In studying the quantity $\phi_c/T_c$ one should in principal take care
to distinguish between various definitions of the critical
temperature.  The $T_c$ relevant for baryogenesis is the temperature
when tunneling from the false vacuum to the true vacuum takes place,
after which the phase transition quickly completes.  To determine it
precisely, one demands that the tunneling probability becomes
sufficiently large compared to the Hubble expansion rate, which
requires solving for the bubble configurations and computing their
action \cite{DLHLL}.  In the standard model, rather close upper and
lower bounds on $T_c$ can be obtained because it is always above the
temperature $T_0$ where $V''(0, T_0) = 0$, and below the temperature
$T_1$ when the nontrivial minimum $v(T)$ in the potential becomes
degenerate with the minimum at the origin, and these temperatures are
all within a few tenths of a percent of each other.  However in
extended models it might be important to make the distinction between
the various critical temperatures because they can differ more
substantially, and also because $v(T)/T$ can change rather quickly near
the critical temperature.   In the following we compare the results of
using either definition of the critical temperature and find some
noticeable differences, particularly for small values of the Higgs mass
parameter ${\mu_3^2}$.

Because the masses of the heavy Higgs bosons ($A^0$, $H^0$ and $H^\pm$)
have such a similar form in eq.~(\ref{Higgsmasses}), we have simplified
our scan of the parameter space by taking them all to be equal to each
other.  Figure 3 shows the contours of constant $\phi/T$ in the plane
of $m_{A^0}=m_{H^0}=m_{H^\pm}$ versus $m_{h^0}$, using three different
methods.  The first column uses $T_0$ as the definition of the critical
temperature, and the Arnold-Espinosa method of ring-improvement.  The
second column is the same except using $T_1$ for the critical
temperature.  The third column also uses $T_1$, but the Parwani
ring-improvement prescription.  Each row is for a separate value of the
potential parameter ${\mu_3^2}$, ranging between 0 and $6\times 10^{4}$
GeV$^2$.

\begin{figure}
\vspace{-1in}
\hbox{\hspace{.75in}$T_0$\hspace{.9in}Arnold-Espinosa
\hspace{.6in} Parwani \hspace{1.1in}$\epsilon$  }
\vspace{0.1in}
\hbox{\hspace{-0.8in}$m_{A^0,H^0,H^\pm}\!\!\uparrow$}
\vspace{0.3 in}\hbox{\hspace{-.5in}${\mu_3^2}=0$}
\vspace{1.4in}\hbox{\hspace{-.5in}${\mu_3^2}=3$} 
\vspace{1.4in}\hbox{\hspace{-.5in}${\mu_3^2}=6$}
\vspace{0.7in}\hbox{\hspace{0.2in}$m_{h^0}\rightarrow$} 
\vspace{-4.8in}
\epsfbox{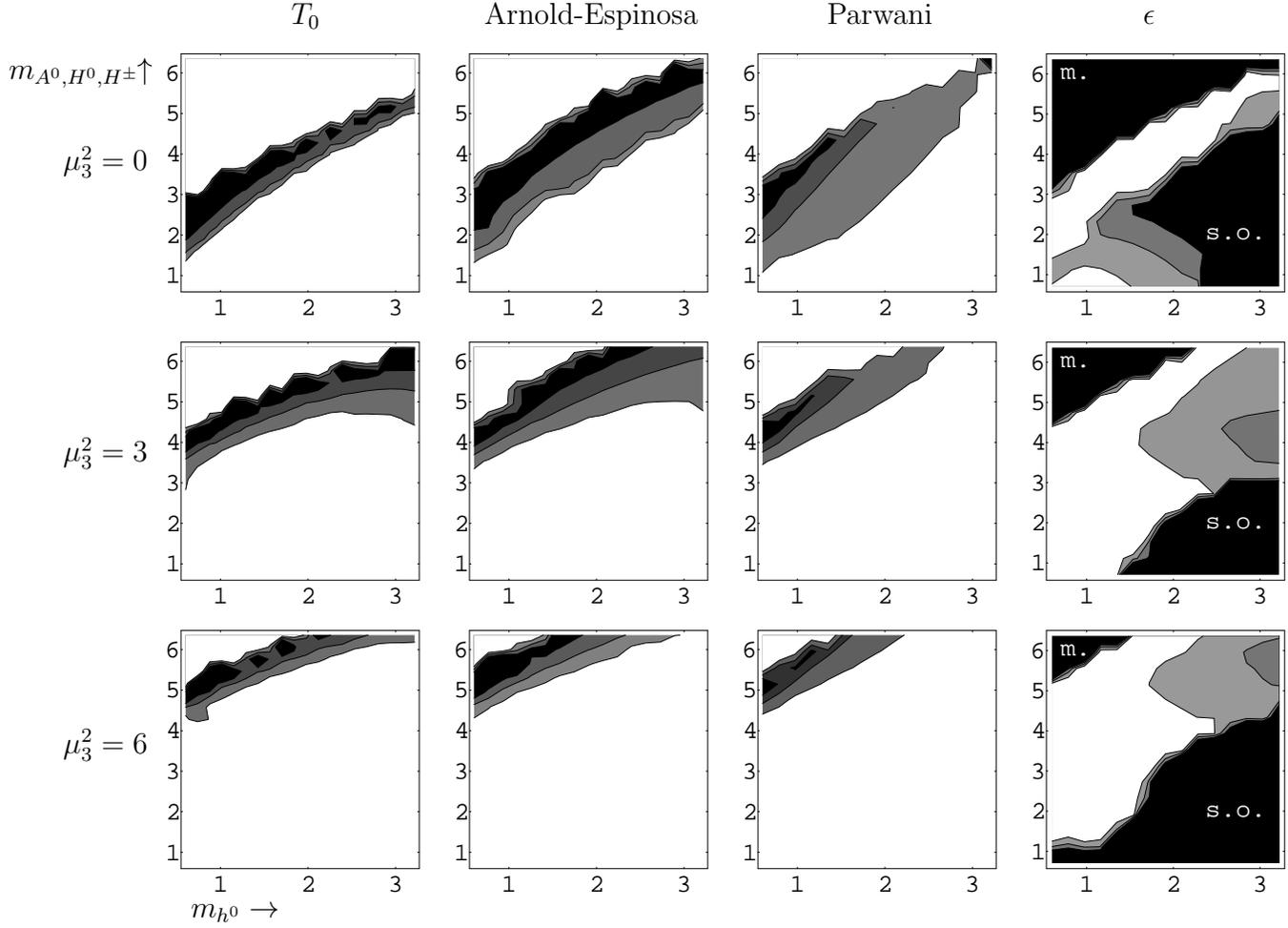}
\vspace{0.2in}
\caption{Contours of constant $\phi_c/T_c$ (first three columns) and
the perturbative expansion parameter $\epsilon$ (defined in
eq.~(\ref{epseq})) for three values of ${\mu_3^2}$, in the plane of
$m_{A^0}=m_{H^0}=m_{H^\pm}$ versus $m_{h^0}$.  The first three columns
use different approximations for the critical temperature or the
effective potential, described in the text.  Units are 100 GeV for
masses, (100 GeV)$^2$ for ${\mu_3^2}$.  Increasing values are represented
by darker shades, with white being $<1$ and increasing in steps of
$1$.  The regions labeled by ``m." in the last column are where the
false vacuum is metastable, and those labeled ``s.o." are where the
phase transition is second order, as illustrated in figure 4.  In
neither case is there a first order transition.  The values of $\epsilon$
are computed using the boson masses corresponding to the second
column.}
\label{fig:results}
\end{figure}

The regions in white are those with $\phi_c/T_c < 1$, and therefore
indicate the parameters for which electroweak baryogenesis would not
work in this model.  In all three methods one sees that for a given
value of $m_{h^0}$, there exists a range of the other masses such that
$\phi_c/T_c > 1$.  When the other masses are greater than this range,
it turns out that the nontrivial minimum is higher than the one at
$\phi=0$ even at zero temperature, meaning that the electroweak
symmetry breaking vacuum is metastable and there is no phase
transition, unless supercooling into the metastable minimum occurs.  On
the other hand, when they lie below the allowed bands, the transition
becomes second order, because there is never any barrier separating the
minimum at $\phi=0$ from the symmetry-breaking minimum.  In either case
we define $\phi_c/T_c$ to be zero.  Such examples are illustrated in
figure 4.

\begin{figure}
\epsfbox{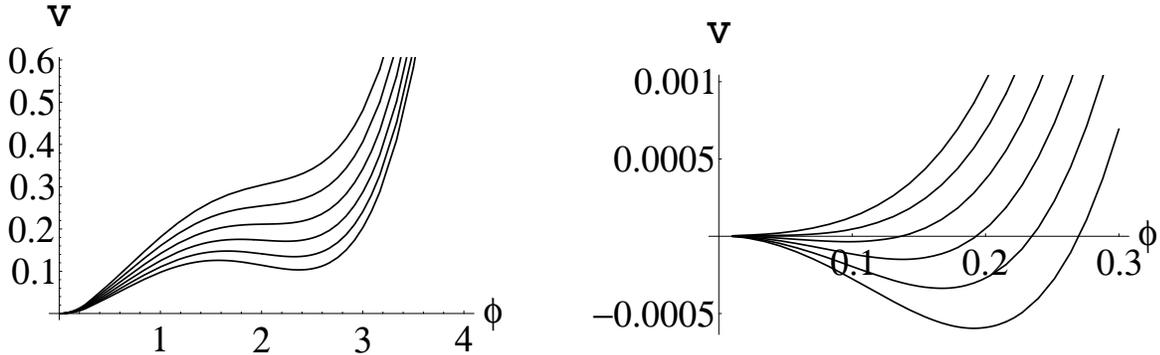}
\caption{Examples of the evolution of the potential with temperature
which illustrate the absence of a first order phase transition.  In the
first one the false vacuum is metastable, and in the second the phase
transition is second order.  The parameter values are $m_{h^0}=60$ GeV,
other masses $= 325 $ GeV, ${\mu_3^2} = 0$ for the first one, and
$m_{h^0}=300$ GeV, other masses $= 250 $ GeV, ${\mu_3^2} = 0$ for the
second.}
\label{fig:failures}
\end{figure}

However the three methods of constructing the finite-temperature
potential differ as to the precise values of $\phi_c/T_c$ when
${\mu_3^2}$ is small, with the Parwani ring-improvement giving larger
values than that of Arnold and Espinosa.  For large ${\mu_3^2}$, all
three give similar results.  Where the two methods of resummation
differ is indicative of where perturbation theory may be unreliable.
There is another curious phenomenon for small but nonzero values of
${\mu_3^2}$, for example ${\mu_3^2} = (30$ GeV)$^2$, which is not shown
in figure 3:  the region allowed by baryogenesis appears to be
significantly enlarged in the first two methods ($T_0$ and
Arnold-Espinosa), though not in the third.  Examination of the
potential shows that the region of enlargement is always accompanied by
the existence of double minima, as in figure 2.  Therefore the
augmentation of the baryogenesis-allowed parameter space in these cases
is spurious.  This is one possible explanation for our failure to
corroborate the claim of ref.~\cite{DFJM}, which says that increasing
${\mu_3^2}$ generally has a favorable effect on the strength of the
phase transition.

In addition to comparing different methods of ring-improvement, we can
also quantify the convergence of the perturbative expansion in an
independent way, by trying to estimate what is the true expansion
parameter for the finite-temperature theory, near the critical value of
$\phi$. Let us first review how the perturbative power counting
argument goes in the standard model. The cost of adding one additional
thermal loop of $W$ bosons to an arbitrary Feynman diagram contributing
to the free energy is
\beq
{g^2 T_c\over (2\pi)^3}\int {d^3p\over (p^2 + m^2_W(\phi_c))^{2}}
        \propto {g^2 T_c\over m_W(\phi_c)} 
        = {6\pi\, m_W\, m^2_{h^0}\over 2 m^3_W + m^3_Z} 
        \cong  \left({m_{h^0}\over 35 {\rm\ GeV}}\right)^2,
\label{expar}
\eeq
where we used $m_W(\phi_c) = g\phi_c/2$, $m^2_{h^0} = 2\lambda_{\rm
eff} v^2$, and the standard model value for critical ratio
$\phi_c/T_c\cong \phi_0/T_0 = (2 m^3_W + m^3_Z)/(2\pi\lambda v^3)$.  In
ref.~\cite{KLRS1} it was shown that the true loop expansion parameter
is around $0.4$ for a Higgs mass of 35 GeV, while ref.~\cite{AE} finds
that $\phi_c/T_c$ changes by 30\% in going from one loop to two loops
when $m_{h^0}$ is near 35 GeV.  In a two-Higgs doublet model, it is
no longer the $W$ boson which gives the dominant corrections to the
potential, but rather the new Higgs bosons.  Thus the expansion
parameter will be $\lambda_i T_c/m^2_i(\phi_c)$ instead of
(\ref{expar}), where $\lambda_i$ and $m_i$ are any of the quartic
couplings and corresponding mass eigenvalues, respectively, of the new
Higgs sector.  The couplings of the mass eigenstates will be linear
combinations of the couplings in the Lagrangian, $\lambda_1$, $h_1$,
$h_2$ and $h_3$, which are given by eq.~(\ref{lameq}) and the relations
\cite{CKV}
\beqa
h_1 &=& 2\lambda(2{\mu_3^2} - 2m^2_{\pm} +m^2_{A^0})/m^2_{h^0};\nonumber\\
h_2 &=& 2\lambda(m^2_{h^0}-2{\mu_3^2} + 2m^2_{\pm}
-m^2_{H^0})/m^2_{h^0};\nonumber\\
h_3 &=& \lambda(2{\mu_3^2} - m^2_{A^0})/m^2_{h^0}.
\label{couplings}
\eeqa
Comparing with the standard model, it thus seems reasonable and
conservative to require that
\beq
\epsilon\equiv {\max(\lambda_1, h_i) T_c\over \min(|m_i(\phi_c)|)} \lsim 1
\label{epseq}
\eeq
in order to have some confidence that the perturbative expansion is not
out of control due to the infrared problem.  In the last column of
figure 3 we show the contours of $\epsilon$, using as inputs the
thermal masses at $\phi_c$ and $T_c$  corresponding to the third
column.  It is interesting to see that $\epsilon$ and $\phi_c/T_c$ tend
to be anticorrelated: where one is big the other is small.  This is
good news as concerns the trustworthiness of the interesting regions
where $\phi_c/T_c>1$; it means we expect the value of $\phi_c/T_c$
predicted by the 1-loop result to be rather stable against changes due to
higher order contributions.

\subsection{Supersymmetry}

\begin{figure}
\epsfxsize 6.5in\epsfbox{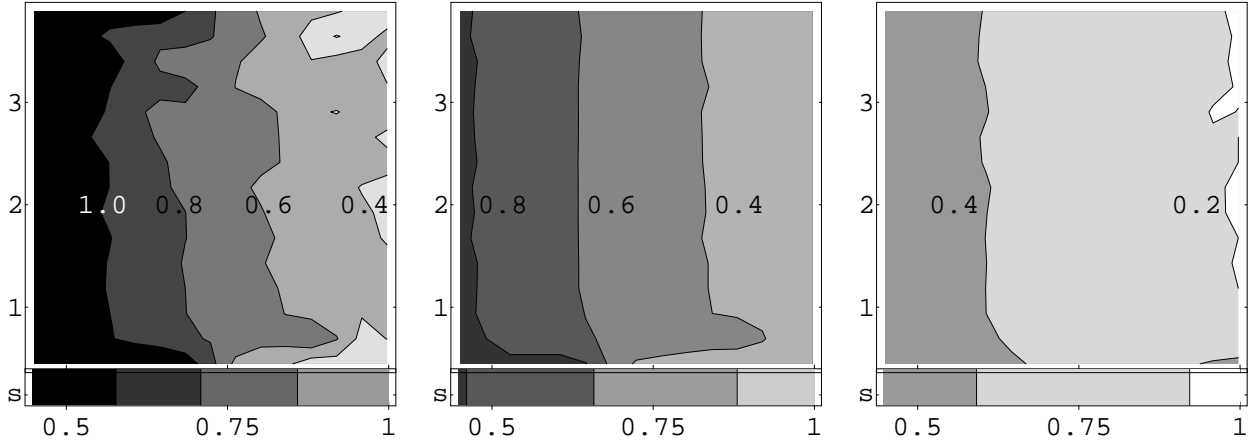}
\caption{Contours of $\phi_c/T_c$ in the plane of $m_{A^0}$ versus
$m_{h^0}$, using the same three respective methods as in figure 3,
setting the other Higgs boson masses to their tree-level values in the
minimal supersymmetric standard model. The horizontal strip at the bottom,
labeled ``s,'' shows the corresponding values in the standard model.}
\label{fig:susy}
\end{figure}

The model we chose to study is not quite a realistic limit of the
minimal supersymmetric standard model (MSSM), even in the case that all
the superpartners are much heavier than the Higgs bosons and therefore
decouple.  This is because to be consistent with the assumption that
the symmetry breaks in the direction where the two Higgs fields have
equal VEV's, we assumed that the top quark coupled equally to each,
which is not possible in supersymmetry.  Nevertheless one might trust
the model to give a qualitative indication of the effects of the additional
Higgs bosons on the phase transition.  In the limit of $\tan\beta = 1$
to which we restricted ourselves in this study, the tree-level masses
of the Higgs bosons are related according to \cite{HHG}
\beq
	m^2_{h^0}=0;\quad m^2_{H^0} = m^2_{A^0} + m^2_Z;\quad
	m^2_{H^\pm} = m^2_{A^0} + m^2_W,
\eeq
and the parameter ${\mu_3^2}=m^2_{A^0}/2$, as can be seen from
eq.~(\ref{couplings}) using the fact that $h_3=0$ in supersymmetry.  Of
course the $h^0$ boson is not really massless, but it gets a mass from
loop corrections involving the top squarks, of order $\alpha
(m^2_t/m_Z)\ln(\tilde m_t/m_t)$.  Thus we are only interested in rather
light $h^0$ particles.  For the heavier Higgs bosons on the other hand,
the tree level relations should be a good approximation for their
masses, showing that $m_{A^0}$ is the only free parameter.  

In figure 5 we display the contours of $\phi_c/T_c$ using the same
three methods as in figure 3, in the plane of $m_{A^0}$ versus
$m_{h^0}$.  For comparison, the standard model results are shown in the
horizontal strip at the bottom of each graph. In contrast to figure 3,
here we see that the distinction between the more accurate critical
temperature and the naive one, $T_0$, is quite important, and
furthermore the differences between the two methods of ring-improvement
are significant.  The latter observation, comparison with the standard
model results, and the lack of variation of $\phi_c/T_c$ with $m_{A^0}$
show that the extra Higgs bosons are generally playing a small role; it
is the transverse gauge bosons which are most important in determining
the strength of the phase transition.  The perturbative expansion
parameter corresponding to their contributions, eq.~(\ref{expar}), is
large for $m_{h^0} > 35$ GeV, hence the discrepancy between the two
methods of resummation.  Although there is a hint that very light
masses for the extra Higgs bosons can strengthen the phase transition,
the apparent instability of the perturbation series in this region
renders any such conclusion uncertain.

It is interesting to compare the MSSM case to that shown in figure 3.
In the latter one sees that for a fixed value of ${\mu_3^2}$ it is always
possible to choose the other masses large enough to make $\phi_c/T_c >
1$.  But in the MSSM, there is a special relation between ${\mu_3^2}$ and
the new Higgs boson masses which prevents them from enhancing the first
order nature of the phase transition in the minimal supersymmetric
standard model, even when the bosons become quite heavy.  This is
easily understood in terms of the couplings $h_i$ in
eq.~(\ref{couplings}), which are constrained to be of order $g^2$ in
supersymmetry, whereas in a generic two Higgs doublet model they become
large as the scalar masses increase.

Also interesting is the large difference between using the naive
critical temperature $T_0$ and the more accurate one $T_1$.  The former
gives a substantial overestimate of the strength of the phase
transition.  In ref.~\cite{DFJM}, regions of parameter space consistent
with $\phi_c/T_c>1$ were found in the MSSM with $\tan\beta=1$, however
using $T_0$ as the critical temperature.  Our results indicate that one
should use a more precise definition of $T_c$ before concluding that
the Higgs bosons can have much effect on the phase transition.

\section{Conclusions}

We have investigated the electroweak phase transition in a class of
models containing two Higgs doublets, with a global symmetry that
insures electroweak symmetry breaking along the direction $\Phi_1 =
\Phi_2$.  The results are encouraging from the standpoint of
electroweak  baryogenesis, which requires that $\phi_c/T_c > 1$ to
avoid the washout by sphalerons of the baryons that might be produced.
Even for rather heavy $h^0$ Higgs particles up to 300 GeV, it appears
possible to choose the others sufficiently heavy to make a strongly
first order transition, except perhaps when the parameter ${\mu_3^2}$
from the mass term ${\mu_3^2} \Phi_1^\dagger\Phi_2$ becomes too large.
In supersymmetric models however, ${\mu_3^2}$ is automatically tuned in
such a way as to minimize the impact of the extra bosons on the
strength of the phase transition, no matter how heavy they are.

We tried to make our study as reliable as possible by using certain
improvements: inclusion of Goldstone boson contributions, going beyond
the small $M/T$ expansion, and estimating the size of higher order
perturbative effects by comparing different methods of ring diagram
resummation and by computing the expansion parameter of the finite
temperature theory.  The latter checks give us confidence that our
conclusions are quantitatively correct.  However it is always possible
that perturbation theory fails despite such safeguards.  Recently it
has become possible to start investigating the electroweak phase
transition in theories with new physics using nonperturbative lattice
results, by mapping the theory of interest onto an effective lagrangian
of the same light degrees of freedom as in the standard model
\cite{KLRS2}, which has already been analyzed on the lattice.  Steps in
this direction were recently taken for two Higgs doublet models by
reference \cite{Losada}.  It will be interesting to compare the
nonperturbative approach with that pursued here, to verify more
conclusively that there really are conditions when perturbation theory,
which is easier to apply and more amenable to analytic expressions, can
be trusted to give quantitatively accurate results for the finite
temperature effective potential.

\end{document}